\def\mass#1{${\rm{#1\,M}_\odot}$}
\def\chem#1#2{$\rm{^{#2}\kern-0.8pt#1}$}
\def\mchem#1#2{\rm{^{#2}\kern-0.8pt#1}}
\def\reac#1#2#3#4#5#6{$\rm{\,^{#2}\kern-0.8pt{#1}\,({#3}\,,{#4})\,{}^{#6}\kern-0.8pt{#5}\,}$}
\def\mreac#1#2#3#4#5#6{\rm{\,^{#2}\kern-0.8pt{#1}\,({#3}\,,{#4})\,{}^{#6}\kern-0.8pt{#5}\,}}
\def\betap#1#2#3#4{$\rm{\,^{#2}\kern-0.8pt{#1}\,(\beta^+)\,{}^{#4}\kern-0.8pt{#3}\,}$}
\def\betam#1#2#3#4{$\rm{\,^{#2}\kern-0.8pt{#1}\,(\beta^-)\,{}^{#4}\kern-0.8pt{#3}\,}$}
\def\mbetam#1#2#3#4{\rm{\,^{#2}\kern-0.8pt{#1}\,(\beta^-)\,{}^{#4}\kern-0.8pt{#3}\,}}
\def\reacbp#1#2#3#4#5#6#7#8{$\rm{\,^{#2}\kern-0.8pt{#1}\,({#3}\,,{#4})\,{}^{#6}\kern-0.8pt{#5}\,(\beta^+)\,{}^{#8}\kern-0.8pt{#7}\,}$}
\def\reacbm#1#2#3#4#5#6#7#8{$\rm{\,^{#2}\kern-0.8pt{#1}\,({#3}\,,{#4})\,{}^{#6}\kern-0.8pt{#5}\,(\beta^-)\,{}^{#8}\kern-0.8pt{#7}\,}$}

\def\simgr{\mathbin{\;\raise1pt\hbox{$>$}\kern-8pt\lower3pt\hbox{$\sim$}\;}}
\def\simlr{\mathbin{\;\raise1pt\hbox{$<$}\kern-8pt\lower3pt\hbox{$\sim$}\;}}

\documentstyle[psfig]{l-aa}

\begin{document}

\thesaurus{14.01.1, 14.02.1, 19.10.1, 19.39.1}

\title{The survival of $^{\bf 205}${\bf Pb} in intermediate-mass AGB stars}

\author{N.~Mowlavi\inst{1}\and S.~Goriely\inst{2}\and M.~Arnould\inst{2}}

\offprints{N.~Mowlavi}
 
\institute{Geneva Observatory, CH-1290 Sauverny, Switzerland\and 
Institut d'Astronomie et d'Astrophysique, C.P. 226, Universit\'e
Libre de Bruxelles, Bd. du Triomphe, B-1050 Bruxelles, Belgium}
 
\date{Received date; accepted date}

\maketitle
\markboth{N. Mowlavi et al: The survival of \chem{Pb}{205} in 
intermediate-mass AGB stars}{ }

\begin {abstract}
The now extinct \chem{Pb}{205} is a pure s-process radionuclide
($t_{1/2} = 1.5\ 10^7$ y) of possible 
substantial cosmochemical interest. As a necessary complement to the detailed
theoretical study of the nuclear physics and astrophysics aspects of the
\chem{Pb}{205} -
\chem{Tl}{205} pair carried out by Yokoi et al. (1985), and to the recent calculation
of the \chem{Pb}{205} production in Wolf-Rayet stars by Arnould et al. (1997), this
paper addresses for the first time  in some detail the question of the
survival of this radionuclide in thermally pulsing  AGB stars. This
problem is made difficult by the high sensitivity to temperature and density of the
rates of the weak interaction processes that are able to produce or destroy
\chem{Pb}{205}. In view of this sensitivity, a recourse to detailed stellar models
is mandatory. With the help of some simplifying assumptions concerning in
particular the third dredge-up characteristics, some of which (like its depth) being
considered as free parameters, predictions are made for the
\chem{Pb}{205} contamination of the stellar surface at the end of a pulse-interpulse
cycle following a series of a dozen of pulses in three different intermediate-mass
stars  ($M=$ \mass{3},$Z=$ 0.02;
$M=$ \mass{6},$Z=$ 0.02; $M=$ \mass{3},$Z$= 0.001). It is concluded that the chances
for a significant
\chem{Pb}{205} surface enrichment are likely to increase with $M$ for a
given $Z$, or to increase with decreasing $Z$ for a given $M$. More specifically,
following the considered pulses at least, the enrichment appears to be rather
unlikely in the
\mass{3} star with $Z=$ 0.02, while it seems to be much more probable in the other
two considered stars. It is also speculated that the (\mass{3},$Z=$ 0.02) star could
possibly experience  some \chem{Pb}{205} enrichment following later pulses than the
ones considered in this paper.

\end{abstract}

\section{Introduction}

There is at present ample observational evidence that some meteoritic material
of solar system origin carries the signature of the in-situ decay of by now 
extinct radionuclides with half-lives from about $10^5$ to approximately 
$10^8$ y. These observations have far-reaching implications for our knowledge
of the history of the forming solar system, and provide in particular a key 
information about the time span between the last
nucleosynthetic events that have modified the composition of the solar nebula 
and the formation of the solar system solid bodies. For a recent review, the
reader is referred to e.g. Swindle (1993) and references therein.

Among the extinct radionuclides of potential cosmochemical interest, 
\chem{Pb}{205} ($t_{1/2} = 1.5\,10^7$ y) exhibits the distinctive feature of
being of pure s-process nature. If its signature in meteorites could be
identified, it would thus very usefully complement the information provided
by the other extinct radionuclides. Unfortunately, its abundance at
the start of the solidification process in the early solar system is very
poorly known. In fact, just an upper limit on this quantity is available
(Huey \& Kohman 1972).

A detailed theoretical study of the nuclear physics and astrophysics aspects 
of the \chem{Pb}{205} -- \chem{Tl}{205} pair has been carried out by 
Yokoi et al. (1985, hereafter YTA85). These authors have shown that, in 
contradiction with an earlier estimate (Blake \& Schramm 1979), enough  
\chem{Pb}{205} could be produced in certain s-process conditions
to allow the development of a s-process chronometry. This of course
requires a renewed search for the signature of this radionuclide in
meteorites (a step in this direction has been taken by Chen \& Wasserburg 
1994), as well as a reliable enough prediction of the \chem{Pb}{205} yields
from all possible s-process sites.  

In order to explore the latter question, YTA85 have considered 
simple parametric models that can at best mimic the complex conditions 
prevailing in the stellar interiors. The first quantitative attempt to
evaluate \chem{Pb}{205} stellar yields has been conducted by Arnould
\& Prantzos (1986) on grounds of detailed models of massive mass-losing stars
of the Wolf-Rayet type. Their early predictions, confirmed recently by Arnould
et al. (1997), reinforce the view already expressed by YTA85 that indeed
\chem{Pb}{205} could be a valuable s-process chronometer. 
Asymptotic Giant Branch (AGB) stars, on the other hand, have also been considered
by YTA85 as potential sites for the \chem{Pb}{205}
enrichment of the interstellar medium (see also Wasserburg et al. 1994). 
Their estimate, however, is based
on a very schematic representation of the recurrent thermal pulse -
interpulse cycles characterising these objects. In addition, the very important
question of the survival of \chem{Pb}{205} during the interpulse phases has been
left unanswered, nor  has it been tackled by Wasserburg et al. (1994) either.

The aim of this paper is to fill this gap by following quantitatively the 
decay of
\chem{Pb}{205} in essentially neutron-free locations of AGB stars during their
interpulse phases. These computations are based on the $\beta$-transition rates derived
by YTA85 and on detailed AGB models 
for three different stars ($M=$ \mass{3}, $Z=$ 0.02; $M=$ \mass{6}, $Z=$ 0.02;
and $M=$\mass{3}, $Z=$ 0.001) whose layers experiencing the 
pulse -- interpulse episodes are supposed to
be enriched with s-process products, and in particular with \chem{Pb}{205}.
It is then assumed that the third dredge-up (hereafter 3DUP)
is responsible for the transport of the surviving \chem{Pb}{205} to the 
surface of the AGB stars before being ejected into the interstellar medium by 
their winds.

Some generalities about the AGB features of interest here (pulses and 
s-process nucleosynthesis) are briefly reviewed in Sect.~\ref{Sect:AGBmodels}.
The specific
aspects of the $\beta$-decay of \chem{Pb}{205} in AGB conditions are described
in Sect.~\ref{Sect:AGBmodels:Pb205}, and the prescriptions adopted for
the calculation of its abundance evolution in neutron-free AGB layers are
presented in Sect.~\ref{Sect:AGBmodels:code}. The predictions of the
survival of \chem{Pb}{205} in the three considered model stars are presented
in Sect.~\ref{Sect:predictions} and discussed in Sect.~\ref{Sect:discussion}.
Some conclusions are drawn in Sect.~\ref{Sect:conclusions}. 

\section{The AGB stars}
\label{Sect:AGBmodels}

\begin{figure*}[t]
\psfig{figure=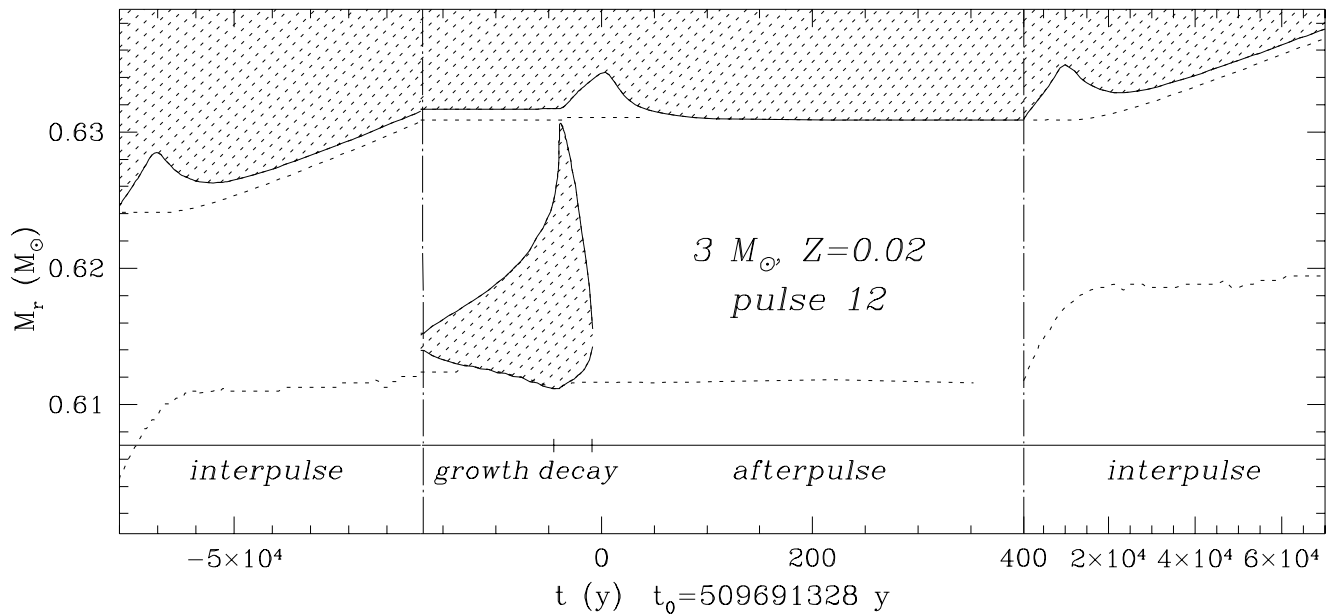}
\caption[]{\label{Fig:phases}
           Structural evolution of the intershell regions of the \mass{3}
           star with $Z = 0.02$ around the 12th pulse. The origin of the
           abscissa is arbitrarily set to time $t_0$, the value of which
           is indicated in the figure. Filled regions correspond to 
           convective zones. The dotted lines identify the
           location of maximum energy production in the H-burning (top) and
           He-burning (bottom) layers. The different phases of an
           instability cycle described in the main text are also indicated
}
\end{figure*}

\subsection{Thermal instabilities}
\label{Sect:AGBmodels:cycle}

The structural evolution of the H- and He-burning shells (denoted hereafter HBS
and HeBS, respectively) during a thermal instability (pulse)
characteristic of the AGB stars is displayed in
Fig.~\ref{Fig:phases}. For the sake of further discussion, four distinct cycle
phases are identified, the main characteristics of which are summarized below 
(more details can be found in Mowlavi 1995):

\smallskip
\noindent Phase 1: {\it The interpulse phase}. Hydrogen burns in a thin layer 
of a few $10^{-4}\,{\rm M_\odot}$ located at mass fractions $M_r$ increasing with 
time. The H-burning products,
including \chem{C}{13} and \chem{N}{14}, are left behind in the
intershell region (layers between the HBS and the HeBS). The HeBS, whose mass
is about $10^{-2}\,{\rm M_\odot}$, is almost
extinct, and the luminosity of the star is essentially provided by the HBS.
As the mass of the He-rich layers increases, a thermal instability is triggered
in the HeBS (Schwarzschild \& H\"arm 1958), leading
to the development of a convective pulse;

\smallskip
\noindent Phase 2: {\it The growth of the convective pulse}. A convective zone
develops just above the location of the maximum energy production of the HeBS, its 
upper boundary coming close to the H-rich layers. The ashes
left behind by the HBS during the interpulse period are engulfed into
the pulse. The pulse duration is about 500 times shorter than the interpulse
one;

\smallskip
\noindent Phase 3: {\it The decay of the pulse}. The energy accumulated in the HeBS
escapes through the now extinct HBS, and the convective pulse recedes;

\smallskip
\noindent Phase 4: {\it The afterpulse phase}. Following the pulse, the structural 
and energetic evolution is complex and dominated by thermal relaxations. The
temperature and density in the He-rich layers first drop as a result of the
expansion of the intershell region, and then increase due to the
structural readjustment. At the same time, the convective envelope
reacts to the thermal relaxations by deepening into
the H-depleted layers, possibly reaching zones containing the ashes of 
the convective pulse. The mixing of these ashes into the envelope and their 
transport to the surface constitute the 3DUP. This dredge-up
is required by the observation of the surface composition peculiarities of AGB
stars, and is predicted by some models after a sufficient number of pulse/interpulse
cycles (Mowlavi 1997). The model pulses dealt with in this paper  (Sect.
5) are not accompanied with the 3DUP, which appears to develop only at
later pulses in the considered stars. However, some
observations suggest that the 3DUP might occur much earlier than
predicted (Mowlavi et al. 1996). In view of the uncertainties still affecting the
precise time of occurrence and extent of the 3DUP, we simply assume in
the following that it indeed takes place at the time of deepening of
the convective envelope in all the pulses to be discussed later. This
time also defines the end of the afterpulse phase, and
correspondingly the start of the following interpulse period.
 
\subsection{The s-process}
\label{Sect:AGBmodels:s-process}

While observation demonstrates that the s-process can develop in AGB stars,
its modelling remains a warmly debated question. This concerns more
specifically the precise mechanism of production and the amount of neutrons
made available for the process. The two classically envisioned  neutron sources,  
\reac{C}{13}{\alpha}{n}{O}{16} and \reac{Ne}{22}{\alpha}{n}{Mg}{25}, have an
efficiency that is expected to depend very much on the considered stars, as well as
on the precise evolutionary phase of a given star. The reliability of the
predictions in this field is drastically limited by astrophysics and nuclear
physics uncertainties (Drotleff et al. 1993, Frost \& Lattanzio 1995).

In the models considered in the present paper, the temperatures limit the
neutron production to \reac{C}{13}{\alpha}{n}{O}{16}. The available
\chem{C}{13} is found to come from the CN cycle exclusively. This amount is 
well known to be insufficient for producing the neutrons required for the
development of a full s-process. This problem classically encountered in the
modelling of the s-process in AGB stars is circumvented by assuming that
some protons from the envelope are brought  ``semi-convectively" into the
underlying \chem{C}{12}-rich layers during the afterpulse phase  (Iben \& Renzini
1982). Recent evolutionary calculations (Herwig et al. 1997) predict similar
mixing to occur by convective overshoot at the base of the envelope.
 In both cases, after reignition of the HBS in the 
early interpulse phase, some \chem{C}{13} could be produced by
\reac{C}{12}{p}{\gamma}{C}{13}. Part at least of this \chem{C}{13} of ``primary''
nature  could be subjected to
\reac{C}{13}{\alpha}{n}{O}{16} either during the interpulse phase or later after
ingestion in the subsequent pulse. These two burning modes will be referred to in the
following as the ``Interpulse Scenario (IS)'' and the ``C Pulse Scenario (CPS)'',
respectively. Of course, the s-process abundances emerging from a pulse may
result from the complementary and sequential action of the IS and of the CPS.
We define this scenario as the IS+CPS. 

Even if the 
\reac{C}{13}{\alpha}{n}{O}{16} neutron source is currently believed to be responsible
for the s-process in AGB stars, it remains that the amount of neutrons that can be
produced in such a way is still highly uncertain, and its dependence on stellar mass
and metallicity is unknown. 
 
\section{The production and destruction of {\bf \chem{Pb}{205}}}
\label{Sect:AGBmodels:Pb205}

As emphasized in Sect.~1, \chem{Pb}{205} has the distinctive feature of being
of pure s-process origin, and could thus be synthesized in the 
IS and/or CPS. The produced quantities 
depend of course on the amount of available neutrons, which cannot be 
estimated self-consistently and reliably at the present stage 
of development of the models. 

\begin{figure}[t]
\psfig{figure=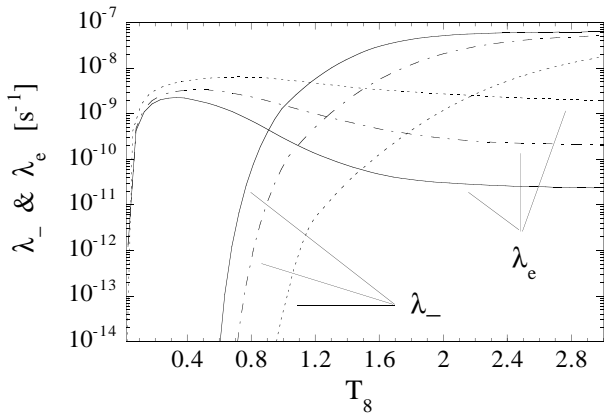,width=9cm}
\caption[]{\label{Fig:rates}
YTA85 rates $\lambda_e$ of \chem{Pb}{205} electron capture and $\lambda_-$ of 
\chem{Tl}{205} $\beta$-decay versus temperature for the electron 
number densities $N_{\rm e}$ (in cm$^{-3}$) of $10^{25}$ (full line), $10^{26}$
(dot-dashed line) and $10^{27}$ (dashed line). The uncertainties 
of nuclear origin affecting these predictions are discussed in detail by
YTA85. They concern mainly some experimentally unknown $\beta$-decay matrix
elements, as well as the \chem{Pb}{205} -- \chem{Tl}{205} mass difference
}
\end{figure}

A complementary question, which has not been addressed properly yet, concerns
the survival of \chem{Pb}{205} in the essentially neutron-free layers in which
the radionuclide may reside after its production by the s-process. Obviously, the
level of
\chem{Pb}{205} destruction in such conditions depends on the relative values of
the relevant stellar evolutionary lifetimes and of the timescales of production or
destruction of this radionuclide through weak interaction processes\footnote{In the
considered neutron-free stellar environments, charged particle reactions or
photodisintegrations cannot affect the \chem{Pb}{205} abundances.}. As
demonstrated by YTA85, the latter lifetimes may depend drastically on the stellar
conditions. This makes the question of the survival of \chem{Pb}{205} a
non-trivial problem. More specifically, in a stellar plasma the  thermal
population of low-lying nuclear excited states, ionization, and the possibility
of capture of free electrons can modify substantially the experimentally known
\chem{Pb}{205} half-life, and even open the possibility for the laboratory stable 
\chem{Tl}{205} to become $\beta$-unstable and to transform into \chem{Pb}{205}.
Figure~\ref{Fig:rates} illustrates the impact of the stellar conditions 
on the \chem{Pb}{205} and \chem{Tl}{205} decay rates. Let us just emphasize
that

\smallskip
\noindent (1) the \chem{Pb}{205} $e^-\!$-capture half-life gets much shorter
than its terrestrial value of $1.5\,10^7$ y at temperatures
that are high enough (typically in excess of a few $10^6$~K) for the first 
excited state of
\chem{Pb}{205} (excitation energy of only 2.3 keV) to be significantly 
populated; 

\smallskip
\noindent (2) when its level of ionization is high enough, \chem{Tl}{205} can
transform into \chem{Pb}{205} via bound-state $\beta$-decay, its corresponding
half-life depending very sensitively on both 
temperature and density. Its lifetime is
shorter than the \chem{Pb}{205} one for temperatures $T \simgr 1.3\,10^8$ K
at a density $\rho \approx 300$ gcm$^{-3}$, this critical temperature
increasing with increasing densities.  

On such grounds, YTA85 predict that \chem{Pb}{205} is in danger of being 
transformed into \chem{Tl}{205} by electron captures in neutron-free regions 
at $10^6 \simlr T \simlr 10^8$ K, the level of destruction being highly
sensitive in this temperature range to the free electron number density and to
the evolutionary timescales. In hotter locations, this destruction
is prevented by the reverse transformation \betam{Tl}{205}{Pb}{205}.
Finally, at $T \simlr 10^6$ K, the \chem{Pb}{205} $e^-\!$-capture lifetime gets
longer than the typical interpulse period ($\simlr 10^6$ y) expected in AGB stars,
so that \chem{Pb}{205} has no time to decay in significant amounts before
the occurrence of the next pulse. Of course, these
temperature limits are only indicative since the rates are density dependent.

In view of the above considerations, it appears hopeless to evaluate in a reliable way
the level of destruction of 
\chem{Pb}{205} in stellar neutron-free locations following the CPS or IS
without recourse 
to a self-consistent abundance calculation conducted in the framework of 
AGB models. The first computation of this kind is described below.
 
\section{The stellar models and numerical assumptions}
\label{Sect:AGBmodels:code}

The time variations of the \chem{Pb}{205} abundance is followed during some
characteristic pulse -- interpulse cycles through detailed model calculations of
three AGB stars with different masses and metallicities referred to in the
following as M3Z02 ($M$=\mass{3},$Z$=0.02), M6Z02 ($M$=\mass{6},$Z$=0.02), and
M3Z001 ($M$=\mass{3},$Z$=0.001),  the evolution of these stars being computed
all the way from the pre-main sequence phase. Details about these evolutionary
models , as well as about the computer code being used can be found in Mowlavi
(1995) and Mowlavi et al. (1996). 

These detailed AGB  models fail to predict self-consistently the s-processing and
the 3DUP, so that some assumptions have to be made in order to
compute the \chem{Pb}{205} surface enrichment that may follow from a thermal
pulse phase. More specifically, we suppose that

\noindent (i) all the considered stars can produce \chem{Pb}{205}. This possibility
is substantiated by the fact that the considered model stars are found to produce
neutrons through the CPS or IS+CPS (see Sect. 5);

\noindent (ii) in line with assumption (i), a \chem{Pb}{205} number density equal to
unity is imposed in the whole convective pulse region at the moment of its maximum
extension in mass, considered as the initial time $t_{in}$ in the abundance
calculations reported below. This choice of initial abundances is justified
by the fact that we are not interested here in the {\it absolute} level of 
\chem{Pb}{205} production, but rather in the extent of its survival after its
synthesis;

\noindent (iii) the 3DUP occurs at the time $t_{dredge}$ of
deepest extent of the convective envelope (see Sect.~2.1). It is considered
to mix homogeneously part of the remaining
\chem{Pb}{205} into the whole convective envelope up to the surface. The fraction of
\chem{Pb}{205} that is effectively transported to the surface depends
on the depth of the 3DUP, taken here as a free parameter. In
addition, the mixing is assumed to take place during a time interval that is short
compared to the lifetimes of the
\chem{Pb}{205} production or destruction mechanisms. These processes are thus
frozen during the 3DUP;

\noindent (iv) after the 3DUP, the \chem{Pb}{205} abundance is followed
in the convective envelope over the time $t_{inter}$ of duration of the interpulse.
In absence of a self-consistent 3DUP model, the physical conditions in the
envelope are assumed to recover their pre-dredge-up values;

\noindent (v) in the $t_{in} \leq t \leq t_{dredge}$ time interval, the layers
loaded with \chem{Pb}{205} following assumptions (i) and (ii) are neutron-free, so
that the
\chem{Pb}{205} abundance is governed by the simultaneous action of just
\chem{Pb}{205}(${\rm e}^-\nu$)\chem{Tl}{205} and
$\mbetam{Tl}{205}{Pb}{205}$. The phase of decay of the modelled pulses (see Sect.~5)
is indeed found to be neutron free, the \chem{C}{13} possibly engulfed in the
pulse convective tongues being fully destroyed at
$t = t_{in}$, while \chem{Ne}{22} cannot burn.

On grounds of the above assumptions, the calculation of the evolution of the
\chem{Pb}{205} abundance profile from
$t_{in}$ up to the end of a given interpulse is then fully coupled to the detailed
modelling of the pulse decay, afterpulse, and interpulse phases. For each stellar
model, we limit ourselves to the consideration of a single pulse-afterpulse-interpulse
cycle identified in Table~1. This simplification is justified by the absence of
self-consistent 3DUP phases in the computed models. It will be dropped in future
models if they succeed to predict such a dredge-up. Even if the number of studied
cycles is limited, they cover a quite large variety of situations
\chem{Pb}{205} can face after its production.

\begin{table}[b]
\caption[]{\label{Table:temperatures}
  Some characteristics at the given pulse number of the three cases for which the
  \chem{Pb}{205} abundance calculations are performed.
  $T_{\rm bp}^{\rm max}$ is the maximum temperature reached at the
  base of the given pulse and $T_{\rm be}$ the temperature at the base of the envelope
  during the following interpulse. The temperatures are expressed in units of
  $10^6$~K. The last column identifies the  scenario assumed to be responsible
for the
\chem{Pb}{205} 
  abundance production (see main text)
  }
\begin{center}
\begin{tabular}{l c c c c}
\hline
\noalign{\smallskip}
 case & pulse & $T_{\rm bp}^{\rm max}$ & $T_{\rm be}$ & scenario\\
\noalign{\smallskip}
\hline
\noalign{\smallskip}
 M3Z02  & 12 & 265 & $\sim  5$ & CPS \\
 M6Z02  & 13 & 285 & $\sim 80$ & IS+CPS \\
 M3Z001 & 14 & 308 & $\sim 30$ & IS+CPS  \\
\noalign{\smallskip}
\hline
\end{tabular}
\end{center}
\end{table}

In radiative neutron-free layers, the \chem{Pb}{205} abundances follow from  
\begin{eqnarray}
 \label{Eq:Pb205}
  \lefteqn{N_{\mchem{Pb}{205}}(t+\Delta t) = } \\
    & & \left( N_{\mchem{Pb}{205}}(t)-N_{205}\frac{\lambda_-}{\lambda}
        \right) \exp^{-\lambda \Delta t} \nonumber
        + N_{205} \frac{\lambda_-}{\lambda},
\end{eqnarray}
with $\lambda = \lambda_- + \lambda_e$, and
\begin{eqnarray}
 \label{Eq:N205}
  N_{205} & = & N_{\mchem{Pb}{205}}(t)+N_{\mchem{Tl}{205}}(t) \nonumber \\
          & = & N_{\mchem{Pb}{205}}(t+\Delta t)+N_{\mchem{Tl}{205}}(t+\Delta t).
\end{eqnarray}
In these equations, $N_i(t)$ is the abundance of nucleus $i$ at time $t$,
$\Delta t$ the time step over which the abundance change is calculated. Following
assumption (ii) of Sect. 4, $N_{\mchem{Pb}{205}}(t_{in}) = 1$ and
$N_{\mchem{Tl}{205}}(t_{in}) = 0$ are adopted. On the other hand, 
$\lambda_-$ and $\lambda_e$ are the values of the \chem{Tl}{205} $\beta$-decay and 
\chem{Pb}{205}
$e^-$-capture rates appropriate for the
temperature, density and composition (and thus electron concentration) of the
considered radiative layer.
Equation~(2) expresses the constancy of the total amount of $A=205$
isobars in absence of any other transformation than the considered 
$\beta$-transitions.

In convective regions, instantaneous mixing is assumed. The evolution of their 
\chem{Pb}{205} content is still given by Eqs.~(1) and (2) in which 
$\lambda_-$ and $\lambda_e$ are interpreted as mass averages of the local
$\beta$-decay and $e^-\!$-capture rates over the whole convective region.
 
\begin{figure}[t]
\psfig{figure=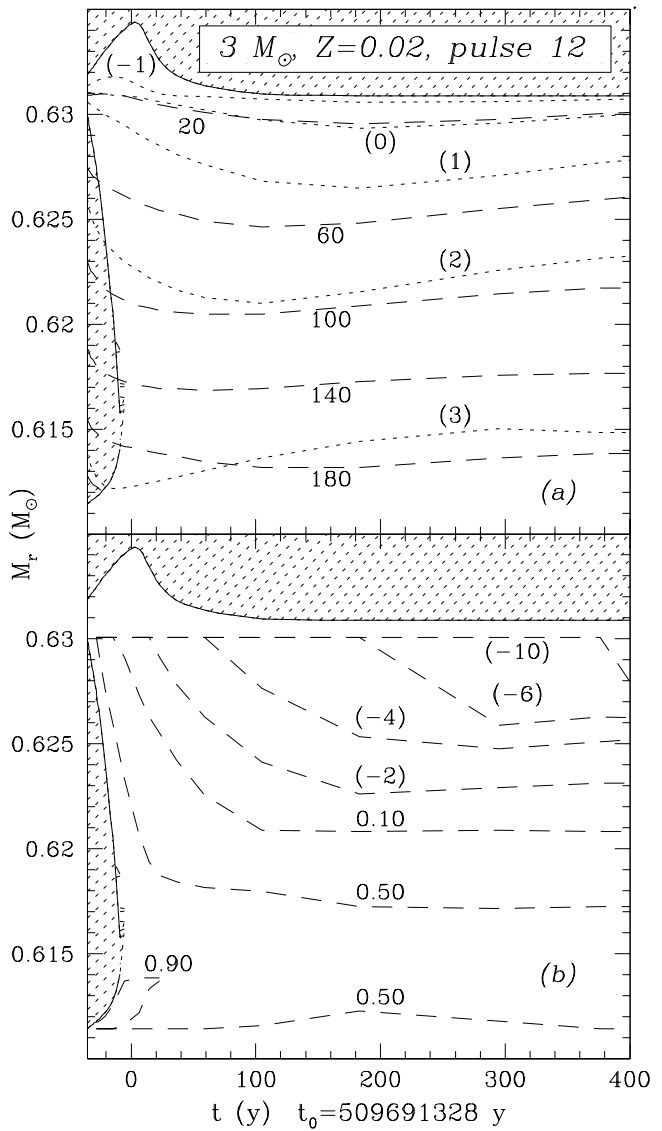}
\caption[]{\label{Fig:s3z02m_cont}
  {\sl(a)} Structural evolution of the intershell regions during the decay of
pulse 12 and during the afterpulse of the M3Z02 star.  Filled areas have the same 
meaning as in Fig.~\ref{Fig:phases}. Long-dashed lines represent locations of
constant temperature, labelled in units of $10^6$~K. Short-dashed
lines indicate locations of constant density (in gcm$^{-3}$). Values in
parentheses indicate logarithms of the density. The origin of time is set at
$t_0$, the value of which is displayed on the abscissa. {\sl(b)} Same as {\sl(a)}
but for contours of  equal \chem{Pb}{205} abundance. Numbers in parentheses refer
to logarithms of the abundances }
\end{figure}

\begin{figure}
\psfig{figure=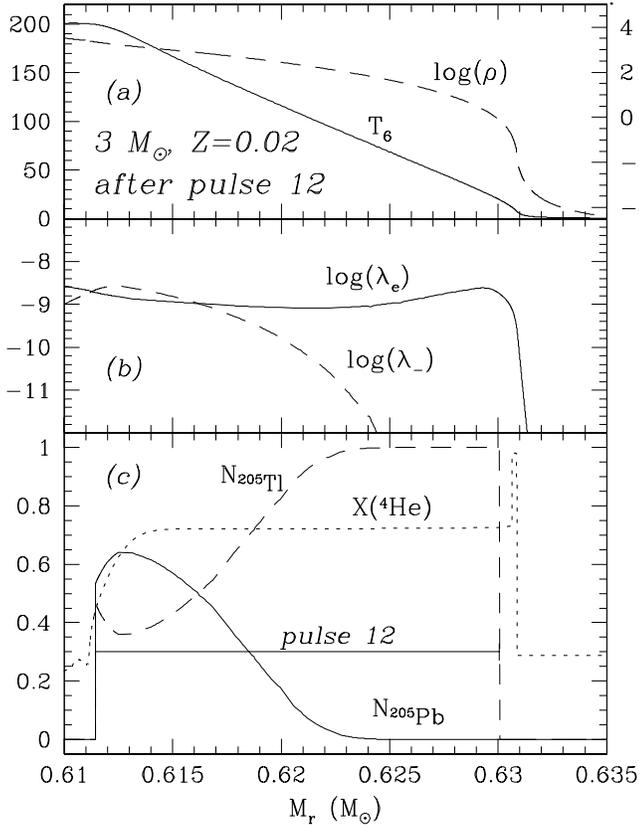}
\caption[]{\label{Fig:s3z02m_PbTl}
  Mass profiles of some quantities in the M3Z02 star at the 
  end of the 12th afterpulse, which occurs at 
  $t=408$~y in the time abscissa of Fig.~\ref{Fig:s3z02m_cont};
  {\sl(a)} Temperatures (in units of $10^6$~K) and densities 
  (in gcm$^{-3}$),
  to be read on the left- and right-hand ordinates, respectively;
  {\sl(b)} \chem{Pb}{205} $e^-$-capture ($\lambda_e$) and 
  \chem{Tl}{205} $\beta^-$-decay ($\lambda_-$) rates, expressed in s$^{-1}$;
  {\sl(c)} \chem{Pb}{205} and \chem{Tl}{205} abundances.
  The helium mass fraction profile (dotted line) helps locating the envelope.
  The extent of pulse 12 is also shown
}
\end{figure}

\section{Model predictions}
\label{Sect:predictions}

\subsection{The M3Z02 case}
\label{Sect:M3Z02}

Up to at least pulse 12, CPS dominates the
\chem{Pb}{205} production. During the decay phase of this pulse, the thermodynamic
conditions in the convective tongue are such that  
\chem{Pb}{205} essentially survives. Though the typical (mass-averaged) lifetime 
of \chem{Pb}{205} is of
the order of the $\approx 30$ y duration of the decay phase, the lifetime
of \chem{Tl}{205} is $\approx 12$ y preventing great
depletions of \chem{Pb}{205}.

\begin{figure}
\psfig{figure=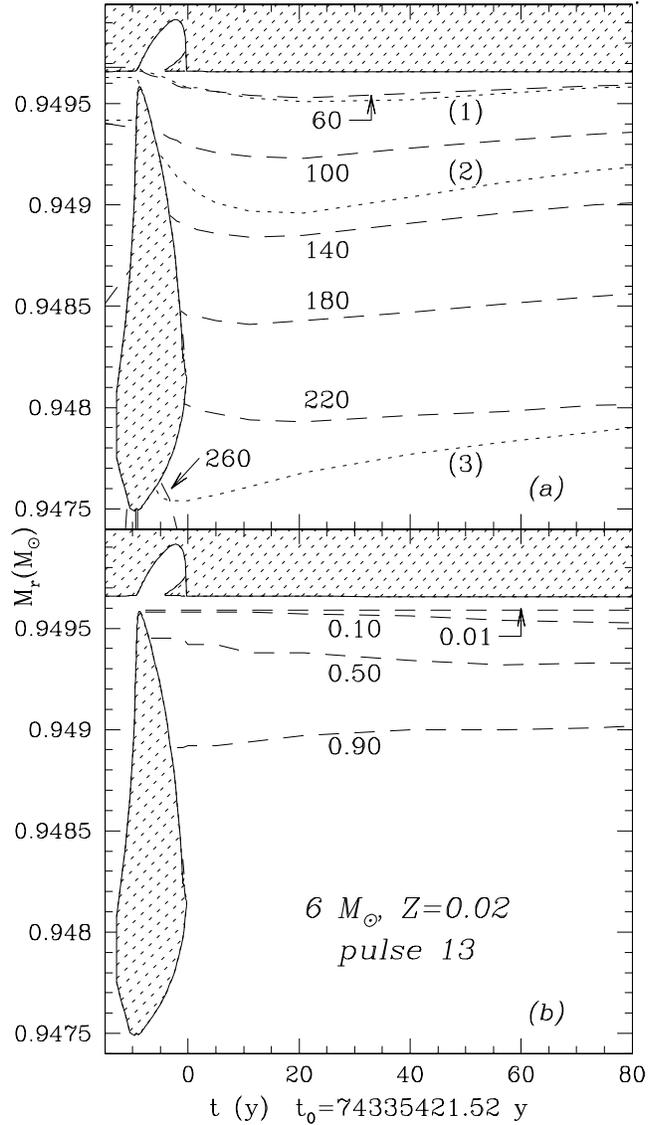}
\caption[]{\label{Fig:s6z02m_cont}
  Same as Fig.~\ref{Fig:s3z02m_cont}, but for pulse 13 of the M6Z02 star
}
\end{figure}

\begin{figure}
\psfig{figure=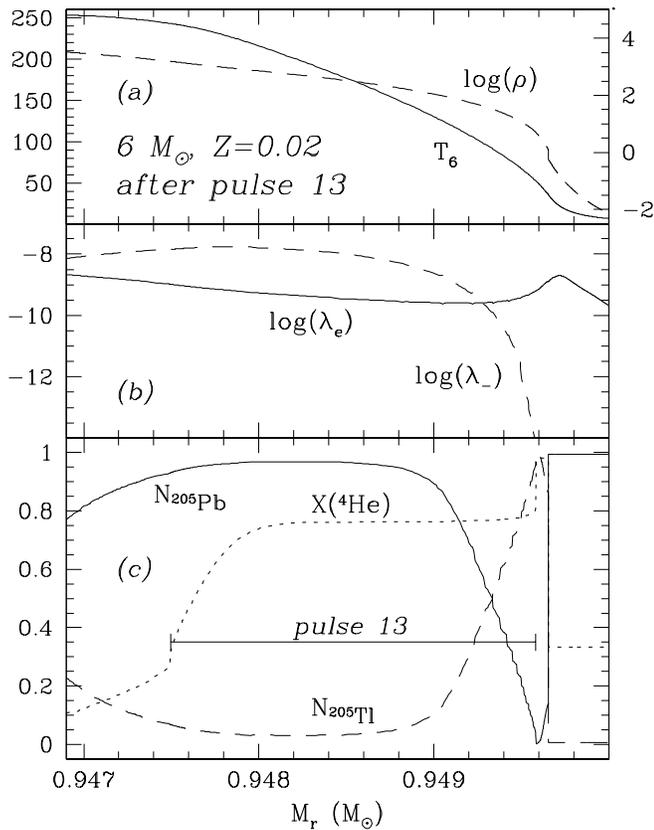}
\caption[]{\label{Fig:s6z02m_PbTlISafter}
  Same as Fig.~\ref{Fig:s3z02m_PbTl}, but at time $t=40$~y (in the scale shown in the
abscissa
  of Fig.~\ref{Fig:s6z02m_cont}) following pulse 13 of the M6Z02 star
}
\end{figure}

In the afterpulse phase, Figs.~\ref{Fig:s3z02m_cont}{\sl(b)} and
\ref{Fig:s3z02m_PbTl}{\sl(c)} demonstrate that the destruction of \chem{Pb}{205}
is severe in layers experiencing  temperatures $T \simlr 10^8$ K, as already
predicted by YTA85 (see also Sect.~\ref{Sect:AGBmodels:Pb205}). This
corresponds to about the upper half of the region formerly covered by pulse 12.
In this mass region, $\lambda_e$ is of the order of $10^{-9}$
s$^{-1}$ [see Fig.~\ref{Fig:s3z02m_PbTl}{\sl(b)}], corresponding to a \chem{Pb}{205} 
lifetime of about 30~y. This is roughly ten times
shorter than the afterpulse duration, so that \chem{Pb}{205} decays 
largely into \chem{Tl}{205}. On the other hand, the reverse transformation is too 
slow to impede this destruction. The situation in this respect is quite 
different in deeper layers vacated by the pulse, where 
$\lambda_-$ can be comparable to $\lambda_e$. In such conditions, the 
destruction of \chem{Pb}{205} is largely avoided, its abundance having time to
reach its local equilibrium value 
$N_{\mchem{Pb}{205}}^{\rm eq} = N_{205} \lambda_-/\lambda$.

The results just described imply that a very deep 3DUP of some $10^{-2}
M_{\odot}$ is required at the end of the afterpulse phase in order to bring some 
\chem{Pb}{205} to the surface of the M3Z02 star. 
Whether or not such a deep 3DUP can operate is still unknown.

\subsection{The M6Z02 case}
\label{Sect:M6Z02}

In contrast to the M3Z02 case, the M6Z02 star at its 13th pulse is the site of
the IS+CPS. Neutrons are not produced anymore at times $t \geq t_{in}$, so that
assumption (v) is fully valid. 

As in the M3Z02 case, almost no \chem{Pb}{205} can
be destroyed during the pulse decay. This results from the fact that, at the
time of maximum extension of pulse 13, \chem{Pb}{205} has an effective (mass-averaged) 
lifetime of the order of 40~y,
while the corresponding \chem{Tl}{205} lifetime is of the order of 3 y. These
lifetimes have to be compared with the approximate 10 y  duration of the pulse
decay.

\begin{figure}
\psfig{figure=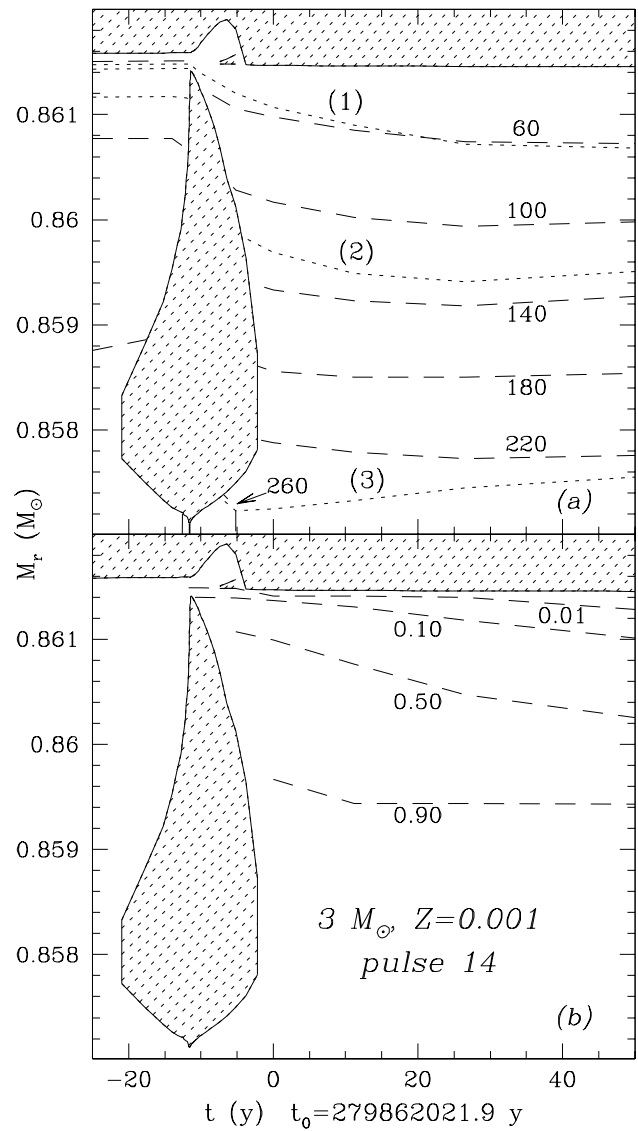}
\caption[]{\label{Fig:s3zs001m_cont}
  Same as Fig.~\ref{Fig:s3z02m_cont}, but for pulse 14 of the M3Z001 star
}
\end{figure}

\begin{figure}
\psfig{figure=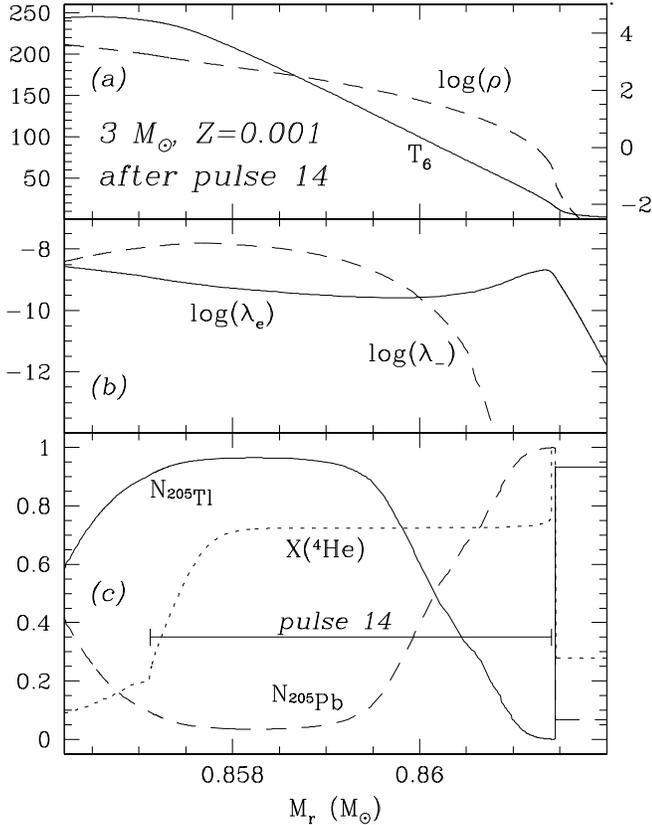}
\caption[]{\label{Fig:s3zs001m_PbTlISafter}
  Same as Fig.~\ref{Fig:s3z02m_PbTl}, but at time $t=60$~y (in the absissa
  of Fig.~\ref{Fig:s3zs001m_cont}) following pulse 14 of the M3Z001 star
}
\end{figure}
   
The M6Z02 afterpulse \chem{Pb}{205} abundance evolution differs in two
ways  from the M3Z02 one. First, the M6Z02 core mass is higher than the M3Z02
one. This implies much higher intershell temperatures in the former case, as
demonstrated  by comparing
Figs.~\ref{Fig:s3z02m_cont}{\sl(a)} and \ref{Fig:s6z02m_cont}{\sl(a)} or
\ref{Fig:s3z02m_PbTl}{\sl(a)} and \ref{Fig:s6z02m_PbTlISafter}{\sl(a)}. Because
of the high sensitivity of $\lambda_-$ to $T$ (Fig.~\ref{Fig:rates}), this
leads to \chem{Tl}{205} $\beta$-decay rates exceeding the \chem{Pb}{205}
$e^-$-capture rates over a large fraction of the intershell region of the M6Z02
star. As a consequence, the \chem{Pb}{205} destruction is prevented over more
than the inner 80\% of the intershell region
[Figs.~\ref{Fig:s6z02m_cont}{\sl(b)} and
\ref{Fig:s6z02m_PbTlISafter}{\sl(c)}]. Second, the response time of the envelope
to the thermal relaxations following the decline of the pulse is about ten times
shorter in the M6Z02 than in the M3Z02 case. This favours further the survival of
\chem{Pb}{205} before the eventual  occurrence of the 3DUP.

In conclusion, large amounts of \chem{Pb}{205} can emerge from massive AGB
stars for moderately deep 3DUP events.

\subsection{The M3Z001 case}
\label{Sect:M3Z001}

There are many similarities between the 14th pulse-interpulse of this star and
the M6Z02 case discussed above. In both stars,  the s-processing is of
the IS+CPS type, and no neutrons can be produced after the maximum extension of the
pulse. In addition, Figs.~\ref{Fig:s3zs001m_cont}{\sl(b)} and
\ref{Fig:s3zs001m_PbTlISafter}{\sl(c)} reveal that much of the initial 
\chem{Pb}{205} survives the afterpulse, and can be brought to the surface by an
eventual 3DUP.

\subsection{The survival of\,\,\chem{Pb}{205} in the envelopes}
\label{Sect:envelope}

If indeed some \chem{Pb}{205} can find its way to the stellar surface, it
remains to be seen if it is able to survive there during the entire duration
$t_{inter}$ of the interpulse phase subsequent to the 3DUP. This is a non
trivial  question as the \chem{Pb}{205} $e^-$-capture and \chem{Tl}{205}
$\beta$-decay rates   depend on the detailed thermal structure of the envelope. 

Table~\ref{Table:envelope} indicates that no significant \chem{Pb}{205}
destruction  occurs in the envelope of the three considered stars, in spite of the 
fact
that the effective \chem {Pb}{205} decay rate in the envelope, obtained by 
mass-averaging 
over all its convective layers, is increased over its
terrestrial value due the high temperatures reached at the
bottom of the envelope. Even so, Table~\ref{Table:envelope} shows that the effective
\chem{Pb}{205} lifetime remains much longer
than the interpulse duration, so that 
\chem{Pb}{205}
has no time to be destroyed significantly in the envelope between two successive
pulses. 

\section{Discussion}
\label{Sect:discussion}

The following remarks are worth emphasizing:

\noindent (i) the stellar models considered in this work lead to some neutron
processing through the CPS or IS+CPS. Interpulses that are hot enough for allowing
the development of a pure IS might well be obtained at later evolutionary phases. In
such conditions, one might wonder if, following its IS production, \chem{Pb}{205}
could not be destroyed before the episode of maximum extension of the following
pulse. This would of course invalidate our assumption of the presence of the
considered radionuclide at this evolutionary phase and, as a consequence, make the
calculations reported in the previous sections meaningless. In fact, we have
performed some numerical tests which suggest that the IS-produced
\chem{Pb}{205}, even if transformed into \chem{Tl}{205} after the IS
neutron-production episode and before the end of the interpulse phase, could be
fully replenished at an early stage of the following pulse growth by the 
fast \betam{Tl}{205}{Pb}{205} transformation taking place at the high temperatures
encountered in the bottom layers of the convective pulse. In such conditions, the
situation would be essentially the same as the one considered in this work, with
\chem{Pb}{205} uniformly distributed in the pulse at its maximum extent. Of course,
more definite conclusions have to await the construction of detailed models leading
to a pure IS production of \chem{Pb}{205};  
 
\begin{table}[b]
\caption[]{\label{Table:envelope}
  Percentage $f^{env}_{Pb,surv}$ of \chem{Pb}{205} in the envelope which survives
 the interpulse of
  the given star. $T_{be}$ is the approximate temperature (in $10^6$~K) at
  the bottom of the envelopes, and $t_{inter}$ the duration (in y) of the
interpulse.
  The last column gives an estimate of the mass-averaged \chem{Pb}{205}
 $e^-$-capture lifetime (in y) in the envelope. 
  }
\begin{center}
\begin{tabular}{lccccc}
\hline
\noalign{\smallskip}
 case & interpulse & $T_{\rm be}$ & $t_{inter}$ & $f^{env}_{Pb,surv}$ & $\tau_e$\\
\noalign{\smallskip}
\hline
\noalign{\smallskip}
 M3Z02  & 12-13 & $\sim  5$ & $\sim 76000$ & 100  & $1.2\;10^7$ \\
 M6Z02  & 12-13 & $\sim 80$ & $\sim  3000$ & 99.4 & $4.8\;10^5$ \\
 M3Z001 & 13-14 & $\sim 30$ & $\sim 10700$ & 93.3 & $1.4\;10^5$ \\
\noalign{\smallskip}
\hline
\end{tabular}
\end{center}
\end{table}
 
\noindent (ii) in the CPS or IS+CPS scenarios encountered in this work, our
ignorance of the precise mass location of the primary \chem{C}{13} responsible for
the neutron release is not likely to influence the fraction of \chem{Pb}{205} that
can survive the thermal pulses. This results from the fact that the
interpulse \chem{C}{13}-loaded layers are efficiently mixed by the time of maximum
extent of the next pulse;

\noindent (iii)  in contrast, the fraction of surviving  
\chem{Pb}{205} drastically depends on the specific
conditions found in the intershell region during the afterpulse phase prior to
the 3DUP;

\noindent (iv) the level of contamination of the stellar surface with
\chem{Pb}{205} is probably quite insensitive to the precise time of occurrence of
the 3DUP. Indeed, Figs.~\ref{Fig:s3z02m_cont}{\sl(b)},
\ref{Fig:s6z02m_cont}{\sl(b)} and \ref{Fig:s3zs001m_cont}{\sl(b)} reveal that,
in regions where \chem{Pb}{205} survives in large amounts, it reaches its
equilibrium value on timescales of the order of the characteristic  
intershell thermal readjustment time, i.e. quite likely well before the occurrence 
of the 3DUP;

\noindent (v) the \chem{Pb}{205} eventually dredged up  to the surface
would not be severely destroyed during the interpulse 
(Sect.~\ref{Sect:envelope}).

\begin{figure}[t]
\psfig{figure=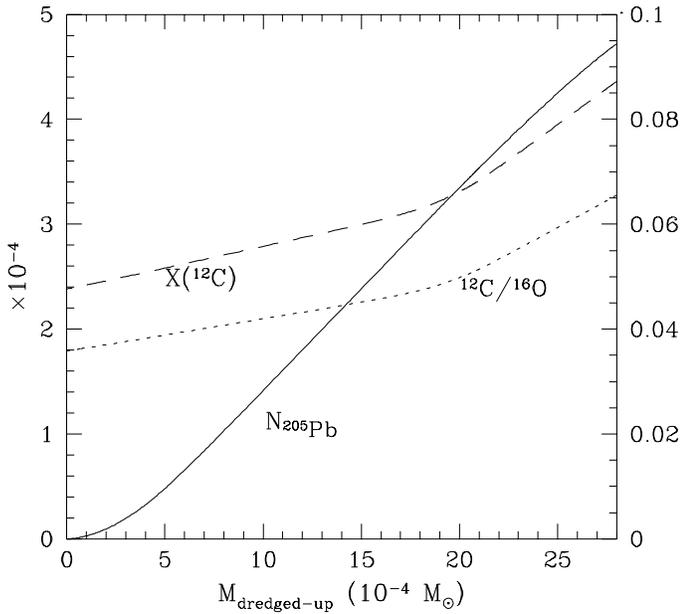}
\caption[]{\label{Fig:dredge-upM}
  Number density of \chem{Pb}{205} and \chem{C}{12} mass fraction (to be read on the left-hand 
  ordinate) and \chem{C}{12}/\chem{O}{16} ratio (to be read on the right-hand
  ordinate)  
  at the surface of the M6Z02 star 40 y after its pulse 13 (see  
  Fig.~\ref{Fig:s6z02m_PbTlISafter}) as a function of the parametrized depth of the
  3DUP within the \chem{C}{12}-rich region left over by pulse 13.
  Note that hot bottom burning at the base of the envelope is responsible
  for the low \chem{C}{12} abundance prior to the 3DUP
}
\end{figure}

Of course, the amount of \chem{Pb}{205} finding its way to the stellar surface
does depend on the depth of the 3DUP. 
Figure~9 illustrates this dependence for the M6Z02 model displayed in
Fig.~\ref{Fig:s6z02m_PbTlISafter}. It is also instructive to present the
surviving
\chem{Pb}{205} at the stellar surface versus the surface \chem{C}{12}/\chem{O}{16}
abundance ratio. Figure~10 shows this correlation after a  single 3DUP following
pulses 12, 13 and 14 in the M3Z02, M6Z02 and M3Z001, respectively. As in
Fig.~9, the depth of the 3DUP is a free parameter. In addition, it
is assumed that the stellar surface is free of \chem{Pb}{205} before the
considered 3DUP, while the \chem{C}{12} and \chem{O}{16} profiles are those
provided by the models.

Figure~10 demonstrates that substantial amounts of \chem{Pb}{205}
could pollute  the stellar surface well before the considered stars become C-type
stars, and that the \chem{Pb}{205} amount increases much more dramatically with
the increased depth of the 3DUP than the corresponding \chem{C}{12} surface
enrichment. Of course, it has to be stressed that the situation displayed in
Fig.~10 is quite artificial, as 
the surface \chem{C}{12} concentration most likely builds up progressively
through successive, and possibly relatively shallow, 3DUPs.  In 
$M \simlr$\mass{3} AGB stars,
in which only very little \chem{Pb}{205} survives in the outer
layers of the intershell region, the succession of pulses could thus lead to an
increase of the surface \chem{C}{12}/\chem{O}{16} ratio without any associated
\chem{Pb}{205} enrichment.
 
\begin{figure}
\psfig{figure=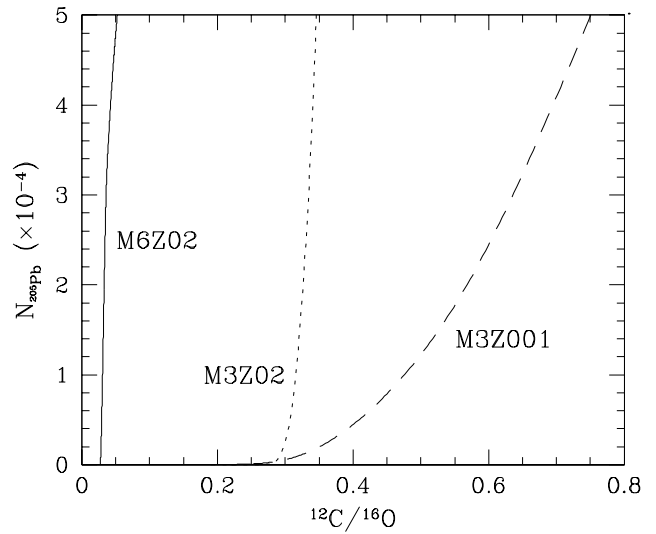}
\caption[]{\label{Fig:dredge-upC}
  Number density of \chem{Pb}{205} 
  as a function of the \chem{C}{12}/\chem{O}{16} ratio at the surface of
  the three stars analysed in the text following the pulse indicated in Table~1.
Each location on the curves corresponds to a given depth of the 3DUP,
which is taken as a free parameter}
\end{figure}

\section{Conclusions}
\label{Sect:conclusions}

This paper presents the first attempt to evaluate in a quantitative way the
possibility of contamination with \chem{Pb}{205} of the surface of thermally pulsing
intermediate-mass AGB stars. We rely on detailed models of 
three different stars ($M=$ \mass{3},$Z=$ 0.02; $M=$ \mass{6},$Z=$ 0.02;
$M=$ \mass{3},$Z=$ 0.001), and on some
simplifying assumptions concerning in particular the 3DUP characteristics, some of
which (like its depth) being considered as free parameters. From the examination in
each of the considered stars of a single pulse-interpulse cycle following a sequence
of a dozen of pulses, we conclude that, during the afterpulse phase, the
\chem{Pb}{205} destruction is prevented in the hottest
($T\simgr 10^8$K) parts of the intershell region. As a direct consequence, the chances
for a significant
\chem{Pb}{205} surface enrichment are likely to increase with $M$ for a
given $Z$, and to increase with decreasing $Z$ for a given $M$. More
specifically, following the considered pulses at least, the enrichment appears to be
rather unlikely in the
\mass{3} star with $Z=$ 0.02, while it seems to be much more probable in the other
two considered stars.  
Of course, the older the AGB star (i.e. the larger the pulse
number), the more \chem{Pb}{205} is likely to survive during the afterpulse, as the
temperatures in the intershell regions increase with the pulse number. It may
thus be that the surface of a \mass{3} star of the M3Z02 type could  be polluted with
a significant \chem{Pb}{205} amount at a more advanced stage of its AGB
evolution, provided of course that the pulse sequence is not too drastically
shortened by severe mass losses. 

Quite clearly, the problem of the \chem{Pb}{205} survival in AGB stars will have to be
revisited more thoroughly once AGB models that predict self-consistently the
development of the s-process and the occurrence of the third dredge-up become
available. In such conditions, it will be meaningful to analyse in detail the
surface \chem{Pb}{205} enrichment resulting from a series of self-consistently
computed pulse-interpulse cycles.

\begin{acknowledgements}
S.~Goriely is F.N.R.S. Senior Research Assistant.
\end{acknowledgements}

\end{document}